\documentclass[notoc,paper,letterpaper]{JHEP3}

\newcommand{\be}{\begin{equation}} 
\newcommand{\ee}{\end{equation}} 
\newcommand{\bea}{\begin{eqnarray}} 
\newcommand{\eea}{\end{eqnarray}}

\newcommand{\bo}{\hbox{1\kern-.23em{\rm l}}}

\def\del{{\partial}}
\def\bar{\overline}
\def\a{\alpha}

\def\ad{{\dot\a}}

\def\th{\theta}
\def\thb{{\bar\th}}
\def\thp{\th^+}
\def\thm{\th^-}
\def\thbp{\thb^+}

\def\qp{q^+}
\def\qm{q^-}

\def\Db{{\bar D}}
\def\Dp{D^+}

\def\Dbp{\Db^+}
\def\Dbm{\Db^-}
\def\Dpp{{D^{++}}}
\def\Dmm{{D^{--}}}

\def\Wb{{\bar W}}

\title{Higher-Derivative Terms in N=2 SUSY Effective 
Actions\footnote{Talk given by GAB at SUSY '03 conference 
(Tucson AZ June 5-10, 2003) and at QTS3 symposium 
(Cincinnati OH September 10-14, 2003).}}

\author{Philip C. Argyres$^{(a)}$, Adel M. Awad$^{(b)}$, Gregory A. Braun$^{(a)}$, 
and F. Paul Esposito$^{(a)}$\\
$^{(a)}$Department of Physics, University of Cincinnati, Cincinnati OH 45221-0011 \\ 
E-mail: \email{argyres,braun,esposito@physics.uc.edu}\\
$^{(b)}$Department of Physics, Faculty of Science,
Ain Shams University, Cairo 11566 Egypt\\
E-mail: \email{awad.adel@usa.net}}

\abstract{
Harmonic superspace can be used to construct higher derivative terms in $\mathcal N=2$ supersymmetric effective actions despite the infinite redundancy in their description due to the infinite number of auxiliary fields.  We are able to write down all of the 3- and 4-derivative terms on the Higgs, Coulomb, and mixed branches, modulo the possible existence of superspace Chern-Simons-like terms, which we discuss.  Many of the terms we find are holomorphic, and at least one is shown to not receive quantum corrections.}

\preprint{UCTP-105-04}

\begin{document}

\section{Introduction}
Higher derivative terms in $\mathcal N=2$ supersymmetric effective actions are of interest because some of them satisfy non-renormalization theorems\cite{ds9705,dg9909}.  Among these higher derivative terms are the Wess-Zumino terms, which have been indirectly shown to exist in $\mathcal N=2$ theories\cite{tz9911,i0001}.  We have performed a systematic exploration of higher derivative terms at the 3- and 4-derivative levels\cite{aabe0306}, which we summarize here.  

\section{Effective Actions}
Low energy effective actions give the approximate behavior of a system at low energies/long wavelengths.  We take the energy scale to be lower than any other scale in the theory, so that only massless particles appear.  Among the bosons we are then left with only $U(1)$ vectors, $A^\mu$, and neutral scalars, $\phi$.  Charged scalars get masses via the Higgs mechanism, and so need not be generically included.

At these low energies, we expand in the number of derivatives, since terms with fewer derivatives will be more relevant at long wavelengths.  To do this expansion, we make use of a derivative dimension counting.  The leading term in the expansion will be of the form $g(\phi) \partial_\mu \partial^\mu \phi$.  Since we want this to be of definite dimension, the function $g$, and therefore $\phi$, must have dimension zero.  Spacetime derivatives have dimension one (since it is they we wish to count), and supersymmetry then demands that vectors are dimensionless while spinors and Grassmann measures $d\theta$ (and equivalently Grassmann derivatives) have dimension $\frac{1}{2}$.  Note that this derivative dimension is not the scaling dimension often used; we introduce
it here just to count derivatives.

\section{${\mathcal N}=2$ Supersymmetry}
These theories have two multiplets.  The hypermultiplet, with superfield $q^+$, and the vector multiplet, with field strength superfield $W$.  In a generic vacuum, the expectation values of the two neutral massless scalar components of $q^+$'s give the 
Higgs branches, while those
of the (neutral) complex scalars in the vector multiplet give the Coulomb branches.  We 
want to find possible higher dimension terms on each of these branches, as well as on the mixed branches, which contain non-zero vevs for both $q^+$ and $W$.

(Note that the field strength $W$ of the Coulomb branch satisfies a constraint.  The unconstrained fields are the vector potential superfields $V^{\pm\pm}$, but are gauge variant and have negative dimension and so are difficult to deal with.  This is because at each derivative level an inifinte number of negative derivative dimension terms may appear.  So, we assume for now that all terms can be written in terms of the gauge invariant field strength $W$.  The possible exception to this, what we call superspace Chern-Simons-like terms, will be discussed later.)

A difficulty is that in any ${\mathcal N}\geq 2$ superspace we need an infinite number of auxilliary fields.  This leads to an infinite number of terms at each derivative level.   Since all the auxiliary fields must integrate out when substituted for using their equations of motion, this infinite number of terms is a unwanted redundancy.

This problem has a solution in the derivative expansion we are using.  When we do this expansion, we are expanding aroung the solution to the 2-derivative, or kinetic term (ignoring Fayet-Illiopoulos terms).  Thus, to look at 3- or 4-derivative (or higher) terms we can use the 2-derivative equations of motion to eliminate the auxiliary terms in favor of propagating fields.  Any corrections to this will be of higher derivative order, and therefore unimportant.  This demonstrates that auxiliary fields remain auxiliary at higher derivative order, and even though they may appear with derivatives, they do not become propagating.  

\section{Harmonic Superspace}
We make use of harmonic superspace\cite{gios01} to organize the infinite number of fields in terms of unconstrained superfields.  Harmonic superspace extends the usual
superspace by an auxiliary space $SU(2)/U(1) \sim S^2$.  This is done by adding auxiliary bosonic $SU(2)$ coordinates $u^\pm$ to the usual superspace, which already has eight fermionic coordinates: $\theta^\pm, \thb^\pm$.  The +/- gives the $U(1)$ charge, so to mod out the $U(1)$ group we look at expressions of definite +/- charge.

In addition, there are derivatives involving only the $u^\pm$: $D^{++}$ and $D^{--}$.  These derivatives and the $u^\pm$ all have derivative dimension zero.
These $u^\pm$ variables can be thought of as living on a 2-sphere.  They are unphysical, and must be integrated out so as not to appear in any physical quantity.

The fields can be written in terms of two different ``halves" of the superspace, each satisfying a constraint.  A chiral superfield satisfies the constraints
$\Db^\pm_\a W = 0$, and is a function of only $x_C$, $\thp$, and $\thm$:
\be
W = W(x_C,\thp,\thm) .
\ee
An analytic superfield satisfies the similar constraints $\Dp_\a \qp = \Dbp_\ad \qp = 0$, and is therefore a function of only $x_A$, $\thp$, and $\thbp$:
\be
q^+ = q^+(x_A,\thp,\thbp) .
\ee
Supersymmetric actions can be made by integrating over one of these halves of superspace:
\be
\int d^2\theta^+\,d^2\thm \; L(\theta^+, \theta^-), \qquad\mbox{or}\qquad
{}\int d^2\thp\,d^2\thbp \; L(\theta^+, \bar\theta^+), \nonumber
\ee
or, over the union of the two, three-fourths of superpace:
\be
\int d^2\thp\,d^2\thm\,d^2\thbp \; L(\theta^+, \theta^-, \thbp), \nonumber
\ee
or finally, over all of superspace:
\be
\int d^8\theta\; L(\theta^+, \theta^-, \bar\theta^+, \bar\theta^-). \nonumber
\ee
Note that we must also, of course, integrate over the unphysical $du$, as well as $d^4x$ to get an action.

By looking at all possible terms of these forms, we are able to get all of the 3- and 4-derivative terms, modulo the possible superspace Chern-Simons-like terms, mentioned earlier and discussed later.

\section{Results}
On the Higgs branches we find only 4-derivative terms, written as an integral of a Grassmann analytic function $B(\qp; u^\pm, \Dpp)$:
\bea\label{higgsres}
S^H_{4} &=&  \int\! du\, d^4x\,
d^2\thp\, d^2\thbp\,\,
\del^\mu \qp \del_\mu \qp B(\qp; u^\pm, \Dpp).
\eea
Each of the $d\theta$ in the measure has derivative dimension $\frac{1}{2}$, and each spacetime derivative has dimension one, and so the term as a whole is dimension four.  Note that this term is holomorphic, in that it depends only on $q^+$, and not its conjugate $q^-$.

On the Coulomb branch, we also find only 4-derivative terms, in the form of an integral over a chiral function ${\mathcal G}(W)$:
\bea\label{coulres}
S^C_{4} &=& \int\! (du)\, d^4x\,
d^2\thp\, d^2\thm \,\,
\del^\mu W \del_\mu W \,{\mathcal G}(W).
\eea
Once again, this terms is holomorphic, since it depends only on $W$, not $\bar W$. 

On the mixed branches we find a number of terms.  First we have a 4-derivative term that is a simple integral over all of superspace:
\bea
S^M_{4c} &= \int\! du\, d^4x\,\,
&d^8\theta\,\,
H(\qp,\qm,W,\Wb; u^\pm,D^{\pm\pm}) .\;
\eea
There is no restriction on $H$; it need not be holomorphic in any of its variables.
Note that there are terms on both the Higgs and Coulomb branches which are integrals over all of superspace, but we consider them to be special case of $S^M_{4c}$.

We also find 4-derivative terms that are written as an integral over three-quarters of harmonic superspace.  Each of these has two spinor derivatives, as well as six $d\theta$'s in the Grassmann measure, for a total derivative dimension of four. 
\bea
S^M_{4a} &= \int\! du\, d^4x\,
&d^2\thp\, d^2\thbp\, d^2\thm\,\, D^+W_a \cdot D^+W_b F^{ab}(\qp,W; u^\pm,D^{++}),\\
S^M_{4b} &= \int\! du\, d^4x\,
&d^2\thp\, d^2\thbp\, d^2\thm\,\Dbm(\Dpp)^n q^+_I \cdot \Dbm(\Dpp)^m q^+_J\nonumber\\
&&G^{IJ}_{nm}(\qp,W;u^\pm,\Dpp).
\eea
Notice that these functions $F$ and $G$ are holomorphic.

Finally, we also find a 3-derivative term,  
\be
S^M_3 = \int\! du\, d^4x\,
d^2\thp\, d^2\thbp\, d^2\thm\,\,
G(\qp,W; u^\pm,D^{++}) .
\ee
This term is important, for it is
the leading contribution (after the kinetic term) to the low energy physics.

These terms may still seem plagued with the infinite redundancy due to the infinite number of possible $\Dpp$ and $\Dmm$ derivatives acting on the $\qp$'s.  This is not, however, the case.  Recall that the second order (2-derivative) equations of motion can be used to eliminate the auxiliary fields at higher derivative order. 
It can be shown\cite{aabe0306} that these second order equations of motion can reduced the number of $D^{\pm\pm}$ derivatives to a finite number, thus making the expressions local in the $u^\pm$'s.  This number of derivatives that it is necessary to include
depends on the specific 2-derivative term in the theory, but is always finite.  For example, if the 2-derivative term for the $q^+$ describes a free hypermultiplet, then only five (plus five more complex conjugate) combinations of $D^{\pm\pm}$ and $q^+$ appear after auxiliary fields are solved for. 

As mentioned, all of these terms except $S^M_{4c}$ are holomorphic functions of their fields.
This limits the types of quantum corrections that may appear.
In $\mathcal{N}=2$ supersymmetric QCD the strong coupling scale $\Lambda$ can be thought of as the lowest component of a field strength superfield $W$.  Thus terms of the form $S^H_{4}$ can receive no $\Lambda$-dependent corrections, since $W$ cannot appear; any such correction would break supersymmetry.  The  other holomorphic terms can only get corrections holomorphic in $W$: one loop and instanton corrections.

It can be shown\cite{aabe0306} that the Wess-Zumino terms are in the $S^M_{4c}$-type mixed branch terms.  The other terms cannot have four antisymmetric spacetime derivatives.
 
\section{Superspace Chern-Simons-like terms}
We must also consider the possibility of terms on the Coulomb branch that cannot be written only in terms of the field strength superfield $W$, but require the use of the potential superfields $V^{\pm \pm}$.
We refer to these as superspace Chern-Simons-like terms in analogy with regular Chern-Simons terms---gauge invariant terms which cannot be written using only the
gauge invaraint field strength $F^{\mu\nu}$, but require also the gauge variant
potential $A^\mu$.  (Though Chern-Simons terms only exist in an odd number of spacetime
dimensions, there is no such restriction for their supersymmetric cousins;  in particular, superspace Chern-Simons-like terms do \emph{not} have regular Chern-Simons
terms in their component expansions.)

Superspace Chern-Simons-like terms are known to exist in $\mathcal{N}=3$ supersymmetric
Yang-Mills in four dimensions\cite{gios01}.  Their existence is still an open
question in $\mathcal{N}=2$ and $\mathcal{N}=1$ superspaces.

\section{Conclusions}

We have shown that harmonic superspace can be used to make a systematic derivative expansion in $\mathcal{N}=2$ supersymmetric theories.  This expansion uses equations of motion from the 2-derivative term to eliminate auxiliary fields at higher derivative order.  We have found all terms at the 3- and 4-derivative level, and have located the Wess-Zumino term.  This classification of terms is modulo the possible presence of superspace Chern-Simons-like terms, the existence of which we are currently investigating.

\section*{Acknowledgments}
This work was supported in part by DOE grant
DOE-FG02-84ER-40153.  G.A.B. was supported in part by a University of 
Cincinnati Research Council's Summer Fellowship.

\end{document}